\documentclass[aps,prl,twocolumn,showpacs,showkeys,superscriptaddress]{revtex4-1}

\usepackage{graphicx}
\usepackage{natbib}
\usepackage{amsmath}

\begin{document}

%\preprint{}
%\newcommand{\sto}{SrTiO$_{3}$}

\title{Epitaxial Growth of Perovskite SrBiO$_3$ Film on SrTiO$_3$ by Oxide Molecular Beam Epitaxy}

\author{Fengmiao Li}
\email{fengmiao.li@qmi.ubc.ca}
\affiliation{Department of Physics and Astronomy, University of British Columbia, Vancouver, British Columbia, V6T 1Z1, Canada}
\affiliation{Stewart Blusson Quantum Matter Institute, University of British Columbia, Vancouver, British Columbia, V6T 1Z4, Canada}

\author{Bruce A. Davidson}
\affiliation{Department of Physics and Astronomy, University of British Columbia, Vancouver, British Columbia, V6T 1Z1, Canada}
\affiliation{Stewart Blusson Quantum Matter Institute, University of British Columbia, Vancouver, British Columbia, V6T 1Z4, Canada}

\author{Ronny Sutarto}
\affiliation{Canadian Light Source, Saskatoon, Saskatchewan S7N 2V3, Canada}

\author{Hyungki Shin}
\affiliation{Department of Physics and Astronomy, University of British Columbia, Vancouver, British Columbia, V6T 1Z1, Canada}
\affiliation{Stewart Blusson Quantum Matter Institute, University of British Columbia, Vancouver, British Columbia, V6T 1Z4, Canada}

\author{Chong Liu}
\affiliation{Department of Physics and Astronomy, University of British Columbia, Vancouver, British Columbia, V6T 1Z1, Canada}
\affiliation{Stewart Blusson Quantum Matter Institute, University of British Columbia, Vancouver, British Columbia, V6T 1Z4, Canada}

\author{Ilya Elfimov}
\affiliation{Department of Physics and Astronomy, University of British Columbia, Vancouver, British Columbia, V6T 1Z1, Canada}
\affiliation{Stewart Blusson Quantum Matter Institute, University of British Columbia, Vancouver, British Columbia, V6T 1Z4, Canada}

\author{Kateryna Foyevtsova}
\affiliation{Department of Physics and Astronomy, University of British Columbia, Vancouver, British Columbia, V6T 1Z1, Canada}
\affiliation{Stewart Blusson Quantum Matter Institute, University of British Columbia, Vancouver, British Columbia, V6T 1Z4, Canada}

\author{Feizhou He}
\affiliation{Canadian Light Source, Saskatoon, Saskatchewan S7N 2V3, Canada}

\author{George A. Sawatzky}
\affiliation{Department of Physics and Astronomy, University of British Columbia, Vancouver, British Columbia, V6T 1Z1, Canada}
\affiliation{Stewart Blusson Quantum Matter Institute, University of British Columbia, Vancouver, British Columbia, V6T 1Z4, Canada}

\author{Ke Zou}
\affiliation{Department of Physics and Astronomy, University of British Columbia, Vancouver, British Columbia, V6T 1Z1, Canada}
\affiliation{Stewart Blusson Quantum Matter Institute, University of British Columbia, Vancouver, British Columbia, V6T 1Z4, Canada}

\begin{abstract}
Hole-doped perovskite bismuthates such as Ba$_{1-x}$K$_x$BiO$_3$ and Sr$_{1-x}$K$_x$BiO$_3$ are well-known bismuth-based oxide high-transition-temperature superconductors. Reported thin bismuthate films show relatively low quality, likely due to their large lattice mismatch with the substrate and a low sticking coefficient of Bi at high temperatures. Here, we report the successful epitaxial thin film growth of the parent compound strontium bismuthate SrBiO$_3$ on SrO-terminated SrTiO$_3$ (001) substrates by molecular beam epitaxy. Two different growth methods, high-temperature co-deposition or recrystallization cycles of low-temperature deposition plus high-temperature annealing, are developed to improve the epitaxial growth. SrBiO$_3$ has a pseudocubic lattice constant $\sim$4.25 \AA, an $\sim$8.8\% lattice mismatch on SrTiO$_3$ substrate, leading to a large strain in the first few unit cells. Films thicker than 6 unit cells prepared by both methods are fully relaxed to bulk lattice constant and have similar quality. Compared to high-temperature co-deposition, the recrystallization method can produce higher quality 1-6 unit cell films that are coherently or partially strained. Photoemission experiments reveal the bonding and antibonding states close to the Fermi level due to Bi and O hybridization, in good agreement with density functional theory calculations. This work provides general guidance to the synthesis of high-quality perovskite bismuthate films.
\end{abstract}

%\pacs{}
%\keywords{Transition metal oxides, Strontium titanate, Oxygen vacancy, and Surface reconstruction}

\maketitle

Hole-doped perovskite bismuthates Ba$_{1-x}$K$_x$BiO$_3$ (x = 0.4) and Sr$_{1-x}$K$_x$BiO$_3$ (x = 0.6) become superconducting below $\sim$30 K and $\sim$12 K, respectively \cite{cava1988superconductivity,kazakov1997discovery}. Their parent compounds BaBiO$_3$ (BBO) and SrBiO$_3$ (SBO) have drawn intense interest \cite{sleight2015bismuthates,cox1976crystal,mattheiss1983electronic,franchini2009polaronic,franchini2010structural,foyevtsova2015hybridization,kim2015suppression,plumb2016momentum,kennedy2006structures,khazraie2018oxygen,balandeh2017experimental,foyevtsov2019structural,naamneh2018cooling,zapf2018domain} owing to the high-temperature superconductivity and also because of their indirect bandgap semiconductor nature rather than metals as would be expected from the half-filled Bi 6s band assuming the formal 4+ valence of Bi. The most common explanation of the nonmetallic behavior invokes the concept of charge disproportionation of Bi$^{4+}$ into the more stable and closed-shell atomic configurations of Bi$^{3+}$(6s$^2$) and Bi$^{5+}$(6s$^0$) \cite{sleight2015bismuthates,cox1976crystal,varma1988missing}. This would inevitably result in a long Bi$^{3+}$-O and a short Bi$^{5+}$-O bond lengths ordered in the simplest conceivable structure with bond length alternation along the three crystallographic axes. Such bond disproportionation is observed in neutron and x-ray diffraction (XRD) \cite{cox1976crystal,kennedy2006structures} and in the resulting band structure \cite{franchini2009polaronic,foyevtsova2015hybridization}. However, x-ray photoemission spectroscopy (XPS)  and density functional theory (DFT) calculations show that the charge difference between the two inequivalent Bi sites is trivial, at most a few tenths of one electron \cite{foyevtsova2015hybridization,plumb2016momentum,wertheim1982electronic,hegde1989electronic,nagoshi1992electronic,shen1989photoemission}. By DFT calculation, Foyevtsova \textit{et al.} recently demonstrates that the bands straddling the Fermi level are in fact strongly hybridized Bi 6s with O 2p molecular orbitals with  A$_{1g}$ symmetry and contain more O 2p character than Bi 6s character, resulting in the bonding states at $\sim$10 eV below the Fermi energy and the unoccupied antibonding states with holes in O 2p just above and electrons just below the Fermi level \cite{foyevtsova2015hybridization}, serving as an alternate explanation for the semiconducting nature. The existence of empty O 2p states just above the chemical potential forming the conduction band was previously mentioned in experiment without reference to its molecular orbital and symmetry character \cite{hegde1989electronic,shen1989photoemission,salem1991determination,kobayashi1999doping}. Further experimental studies are required to compare with the recent DFT calculations. Uncovering the electronic structure in the parent compounds would also facilitate an understanding of the states that lead to superconductivity when doped with holes.

\begin{figure}[t]
\includegraphics[clip,width=3.3 in]{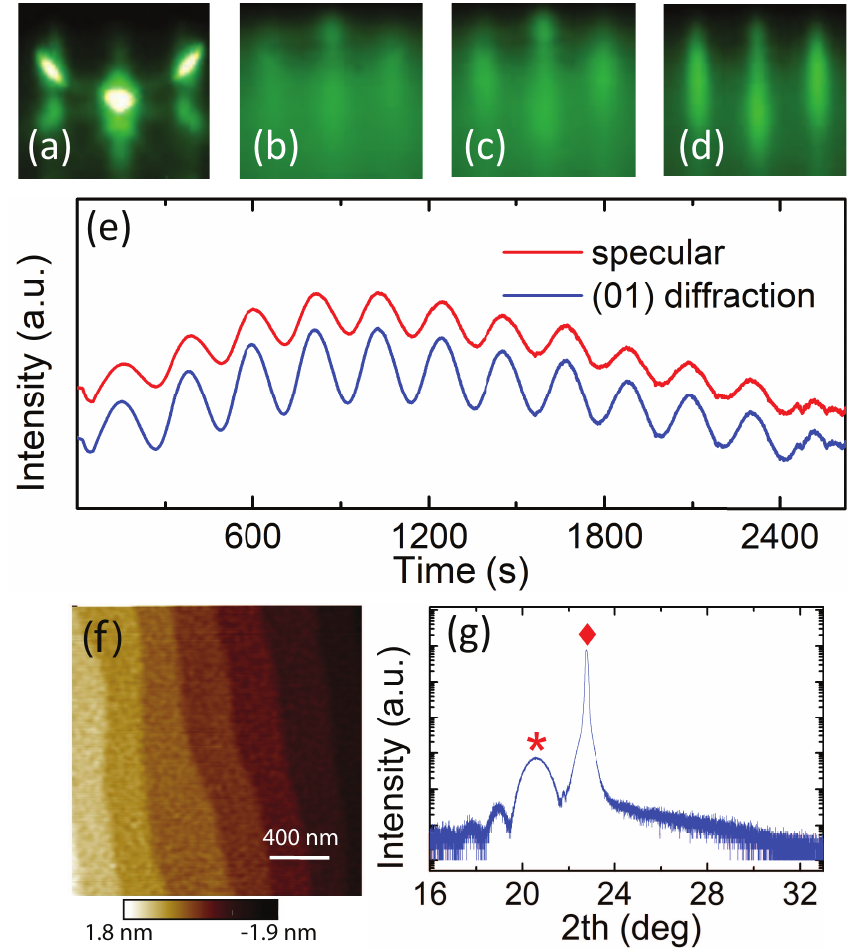}
\caption{\label{fig1-arpes-eels} (a)-(d) [100] direction RHEED pattern of 2~AL SrO (a), 2~UC SBO (b), 4~UC SBO (c) and 20~UC SBO (d) grown by the high-temperature codeposition; (e) RHEED intensity oscillations of the specular and (01) diffraction spot after 6~UC SBO film growth on SrO-terminated STO. (f) AFM image of 20~UC SBO film surface; (g) XRD of the 20~UC SBO film grown on STO substrate. $\ast$ and $\diamond$ label SBO (001) and STO (001) diffraction, respectively.}
\end{figure}

Because of its three-dimensional perovskite structure, cleaving of bulk SBO single-crystals does not yield usefully large, flat areas of known termination. High-quality samples in thin film form are desirable not only because they can facilitate the study of electronic structure using surface science probes like photoemission spectroscopy, but also can accommodate the realization of heterostructures for field-effect doping without the disorder introduced by chemical dopants (\textit{e.g.} K). A number of reports on the growth of undoped or hole-doped BBO films by molecular beam epitaxy (MBE) and pulsed laser deposition have recently appeared \cite{kim2015suppression,plumb2016momentum,zapf2018domain,inumaru2008partial,hellman1989molecular,hellman1990adsorption,gozar2007surface,mijatovic2002growth,lee2016double}. One challenge is to find perovskite substrates to match BBO’s large pseudocubic lattice parameter $\sim$4.35~\AA. SBO has a smaller pseudocubic lattice parameter $\sim$4.25~\AA~resulting in a reasonable lattice match to some large lattice-parameter perovskite substrates such as BaSnO$_3$ ($\sim$4.12~\AA) and LaLuO$_3$ ($\sim$4.18~\AA). However, these two substrates are not commercially available yet. Film quality is also limited by the temperature-sensitive sticking coefficient of Bi, leading to a small growth window to obtain cation stoichiometric films.

In this paper, we report on the successful epitaxial growth of SBO film on SrTiO$_3$ (STO) (001) substrate by oxide MBE. We have used two growth methods, high-temperature co-deposition or recrystallization cycles of low-temperature deposition plus high-temperature annealing, to yield films with correct stoichiometry, high crystalline quality, and atomically flat surfaces. Further, the electronic structure of fully relaxed SBO films is studied using angular integrated photoemission spectroscopy and it agrees well with the density of states (DOS) in DFT.

The films are grown in an oxide MBE chamber with a base pressure better than 1.0~$\times$~10$^{-10}$~Torr. During growth, the film surface is monitored by reflection high-energy electron diffraction (RHEED). The Bi and Sr are evaporated using low-temperature effusion cells and the flux rate is calibrated with a water-cooled quartz crystal microbalance before growth. The Sr flux rate is kept at 4.4~$\times$~10$^{12}$~atoms~$\cdot$~cm$^{-2}$~$\cdot$~s$^{-1}$. Reactive atomic oxygen generated by thermal cracker or plasma is used as oxygen source. As-received TiO$_2$-terminated STO substrates are annealed at $\sim$600~$^{\circ}$C under an O$_2$ pressure of 2.0~$\times$~10$^{-7}$~Torr for 2 hours to remove the surface contamination prior to film growth. Two atomic layer (AL) SrO buffer layers are found necessary to obtain the single-crystal SBO films and are first grown on the TiO$_2$-terminated STO surface with the substrate temperature at 450~$^{\circ}$C in 2.0~$\times$~10$^{-7}$~Torr O$_2$
pressure, for both methods.

The sticking coefficient of Bi depends strongly on the sample temperature and the partial pressure (and species) of oxygen at the sample surface \cite{hellman1989molecular,migita1997self}. To obtain the correct Bi oxidation state and film stoichiometry, tests have been performed by systematically varying growth conditions, including substrate temperature, oxygen pressure, as well as the flux ratio of Sr and Bi. The stoichiometric crystalline films reported here are grown using the optimal conditions as follows. SBO film is grown by co-deposition of Bi and Sr with about 30\% excess Bi at substrate temperature 450~$^{\circ}$C and oxygen pressure 2.0~$\times$~10$^{-7}$~Torr (chamber pressure as measured by an ion gauge; the local pressure close to the substrate will be higher because the end of nozzle used for oxygen delivery is much closer to sample surface). The 30\% excess Bi desorbs from the surface due to its volatility at the growth temperature \cite{hellman1989molecular,migita1997self}. Rutherford backscattering spectrometry (RBS) collected at the Rutgers University Laboratory for Surface Modification, confirms that the Bi/Sr ratio in a film grown on MgO(001) substrate using the same conditions is within $\pm$5\% of 1:1. It is found that a bigger than $\pm$4\% change of the optimum F$_{Bi}$/F$_{Sr}$ ratio leads to an amorphous film or a film surface with islands, with no diffraction spot or multiple diffraction spots along the direction perpendicular to film surface in RHEED. After film growth, the oxygen pressure and thermal cracker/plasma power are kept constant until the sample temperature falls below 100~$^{\circ}$C. Note that films directly grown on the TiO$_2$-terminated surface of STO substrate always show multiple domains with different orientations in our experiments. The need for the SrO layers is likely due to the tendency of Bi to occupy the A-site in perovskite ABO$_3$ structure  and react with the TiO$_2$ of the substrate to form a stable bismuth titanate compound; the SrO layer functions as a diffusion barrier to prevent the formation of this phase.

\begin{figure}[t]
	\includegraphics[clip,width=3.3 in]{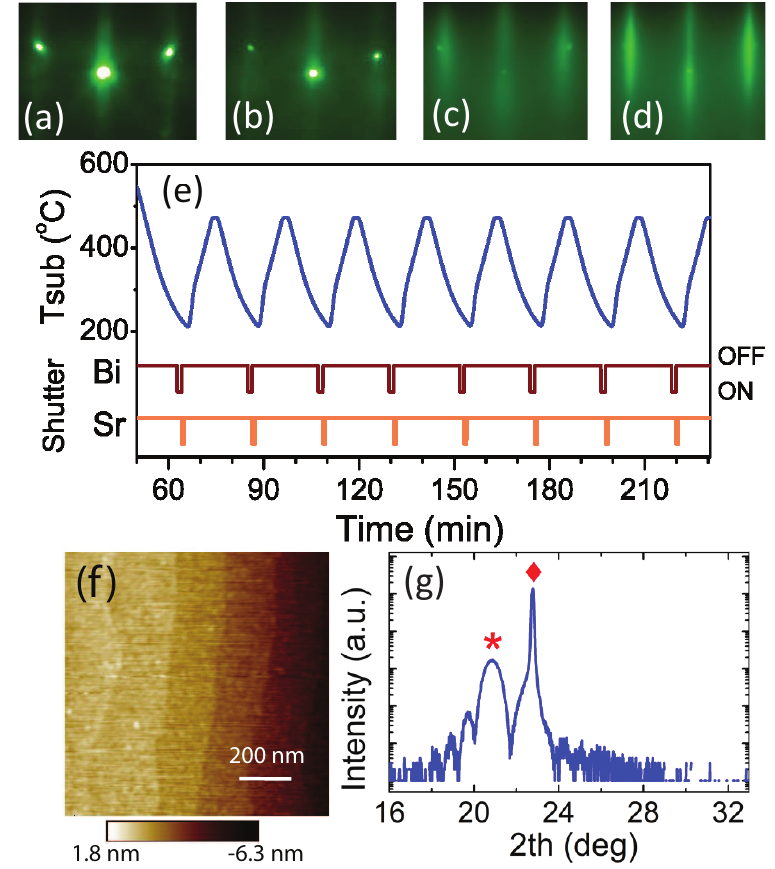}
	\caption{(a)-(d) [100] direction RHEED pattern of 2~AL SrO (a), 1~UC SBO (b), 4~UC SBO (c) and 6~UC SBO (d) grown by the recrystalization method; (e) top panel: substrate temperature cycle during film growth; bottom panel: Sr and Bi shutter open/close status; (f) AFM image of 28~UC SBO film surface; (g) XRD of the 28~UC SBO film grown on STO substrate. $\ast$ and $\diamond$ label SBO (001) and STO (001) diffraction, respectively.}
\end{figure}

Figure~1~(a)-(d) shows RHEED patterns after 2~AL SrO, 2 unit cell (UC) SBO, 6~UC SBO, and 20~UC SBO, respectively, using the high-temperature codeposition method of Bi and Sr. The [100] direction of SBO is parallel to the [100] of the STO substrate, demonstrating cube-on-cube epitaxial growth. The RHEED intensity during the first few layers of growth is first reduced, likely due to the film nonstoichiometry resulted from the extremely low sticking coefficient of bismuth at the begining of growth \cite{zapf2018domain}. The RHEED pattern recovers and its intensity increases after $\sim$6~UC growth indicating improving crystallinity of the SBO phase. As shown in Fig.~1~(d), the 20~UC SBO film has streaky and bright RHEED pattern suggesting an atomically smooth and well-ordered film surface. The separation of the diffracted streaks reflects the in-plane lattice parameter of film in the [100] direction. As shown in Fig.~S1 of Supplemental Material \cite{supplemental}, the pixel spacing of these streaks in the 2~AL SrO buffer layers and the 2~UC film surface are similar to that of the substrate, indicating coherently strained growth. Films more than 6~UC thickness have $\sim$4.25~\AA~ in-plane lattice parameter, as evidenced by $\sim$10\% reduction in separation of diffracted streaks compared to the substrate. These results suggest that the film starts to relax from 2~UC thickness and becomes fully relaxed at $\sim$6~UC thickness. Time-dependent RHEED intensity after 6~UC film growth in Fig.~1(e) shows 12 oscillations revealing the 2-dimensional layer-by-layer growth mode. The atomic force microscope (AFM) of the 20~UC film surface in Fig.~1(f) and Fig. S4 \cite{supplemental} shows atomically flat steps and terraces. X-ray diffraction (XRD) experiment was performed on a Rigaku smartlab pXRD machine and a strong (001) diffraction peak at 2$\theta$~=~20.6$ ^{o} $ is observed in XRD in Fig.~1(g) and Fig. S3 \cite{supplemental}, corresponding to an out-of-plane lattice parameter $\sim$4.25~\AA~ that is consistent with the fully relaxed film seen by RHEED. The narrow rocking curve taken at SBO (001) diffraction with 0.1$ ^{o} $ full width at half maximum (FWHM), and strong finite thickness oscillations in XRD demonstrates the high quality of film.

In high-temperature codepositioon growth, only a small window of Bi/Sr ratio allows us to achieve high-quality stoichiometric films. Intrigued by the growth method of STO on Si \cite{warusawithana2009ferroelectric} and considering the volatility of Bi metal and its oxides at high temperature, we developed a new method for perovskite bismuthate growth, based on cycling between low-temperature deposition and high-temperature crystallization. Low temperature deposition in atomic oxygen enviroment leads to an amorphous SBO layer that avoids Bi re-evaporation from surface, while high-temperature annealing substantially promotes crystallization and improves film quality.

As shown in Fig.~2, in each cycle, atomic layers of Bi and Sr are deposited in atomic oxygen environment to form an amorphous layer at $\sim$200~$^{\circ}$C, followed by an annealing step at 470~$^{\circ}$C. Figure~2(a)-(d) shows the RHEED pattern after 2~AL SrO, 1~UC SBO, 4~UC SBO and 6~UC SBO, respectively. The diffracted streak spacing shows 1-4~UC film is partially strained and becomes fully relaxed at 8~UC as shown in the Fig.~S2 of Supplemental Material \cite{supplemental}, consistent with the obervation in the high-temperature growth method. Note that a much higher quality for films thinner than 6~UC is obtained in this cycling method, because low temperature deposition avoids the re-evaporation of Bi from surface leading to a correct Sr \& Bi ratio in the film \cite{zapf2018domain}. As shown in Fig.~2(f) and Fig.~S5 \cite{supplemental}, steps and terraces are clearly visible in the AFM image of the 28~UC film surface. The high quality of the 28~UC SBO film is demonstrated by the strong (001) diffraction peak in XRD with finite thickness oscillations as shown in Fig.~2(g), showing an out-of-plane lattice constant of $\sim$4.25~\AA. The rocking curve taken at SBO (001) diffraction has 0.08$^{\circ}$ FWHM, similar to that seen in the high-temperature codeposition method. We conclude that the recrystalization growth method works well for SBO growth, and solves the problem of ultrathin film nonstoichiometry due to the low sticking coefficient of Bi at the begining of film growth \cite{zapf2018domain}. This growth method could be applied to growth of other Bi-based oxide systems.

\begin{figure}[t]
	\includegraphics[clip,width=3.3 in]{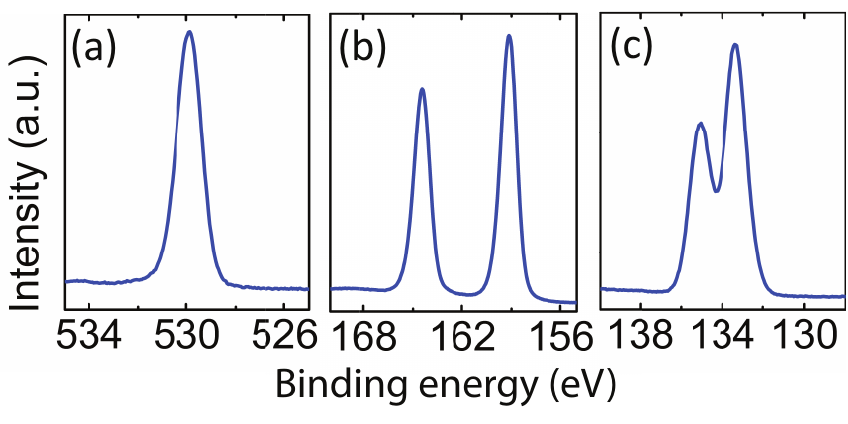}
	\caption{XPS of O1s (a), Bi4f (b) and Sr3d (c) measured on the 20~UC SBO film grown on Nb-doped STO using the high-temperature codeposition method. }
\end{figure} 

The chemical composition and electronic structure of the films are studied using \textit{in situ} x-ray photoemission spectroscopy (XPS) with monochromatized Al \textit{K$_{\alpha}$} (1486.6~eV) x-ray source. Figure~3(a)-(c) shows the O1s, Bi 4f, and Sr 3d core-level spectroscopy of 20~UC SBO film, respectively. The film stoichiometry can be determined from the XPS intensities after taking into account the element sensitivity factor which is calibrated by measuring STO substrate and Bi$_2$O$_3$ film \cite{moulder1995handbook}, and the result shows film stoichiometry close to SBO with a $\sim$$\pm$5\% error bar, consistent with the RBS result. 

As shown in Fig.~3(a), the O 1s spectroscopy with peak at $\sim$529.8~eV binding energy [Fig.~3(a)] is a little asymmetric, indicating the existence of a small ``satellite'' peak at higher binding energy side which may result from the presence of O 2p holes interacting with the core 1s hole in the XPS final state \cite{balandeh2017experimental}. The double peak spliting of Bi 4f [Fig.~3(b)] and Sr 3d [Fig.~3(c)] is due to spin-orbital coupling. As manifested by the single Bi 4f$ _{7/2} $ peak in Fig.~3(b), the Bi in our SBO film has a single valence. However, it is difficult to determine the precise valence state of Bi based on the peak position alone due to the variance of E$ _F $ position in different bismuth oxides, together with the differences in the Madelung potentials and degrees of covalency. The Bi 4f$ _{7/2} $ and O1s spectra show a shift of $\sim$1.2~eV to higher binding energy as compared to that in BBO \cite{nagoshi1992electronic}. This could be explained in part by the larger indirect band gap in SBO, suggesting that the chemical potential is located below the bottom of conduction band \cite{foyevtsova2015hybridization}. Transport measurement shows the film is insulating, indicating the Fermi level located in the band gap, close to the conduction band.

\begin{figure}[t]
	\includegraphics[clip,width=2.7 in]{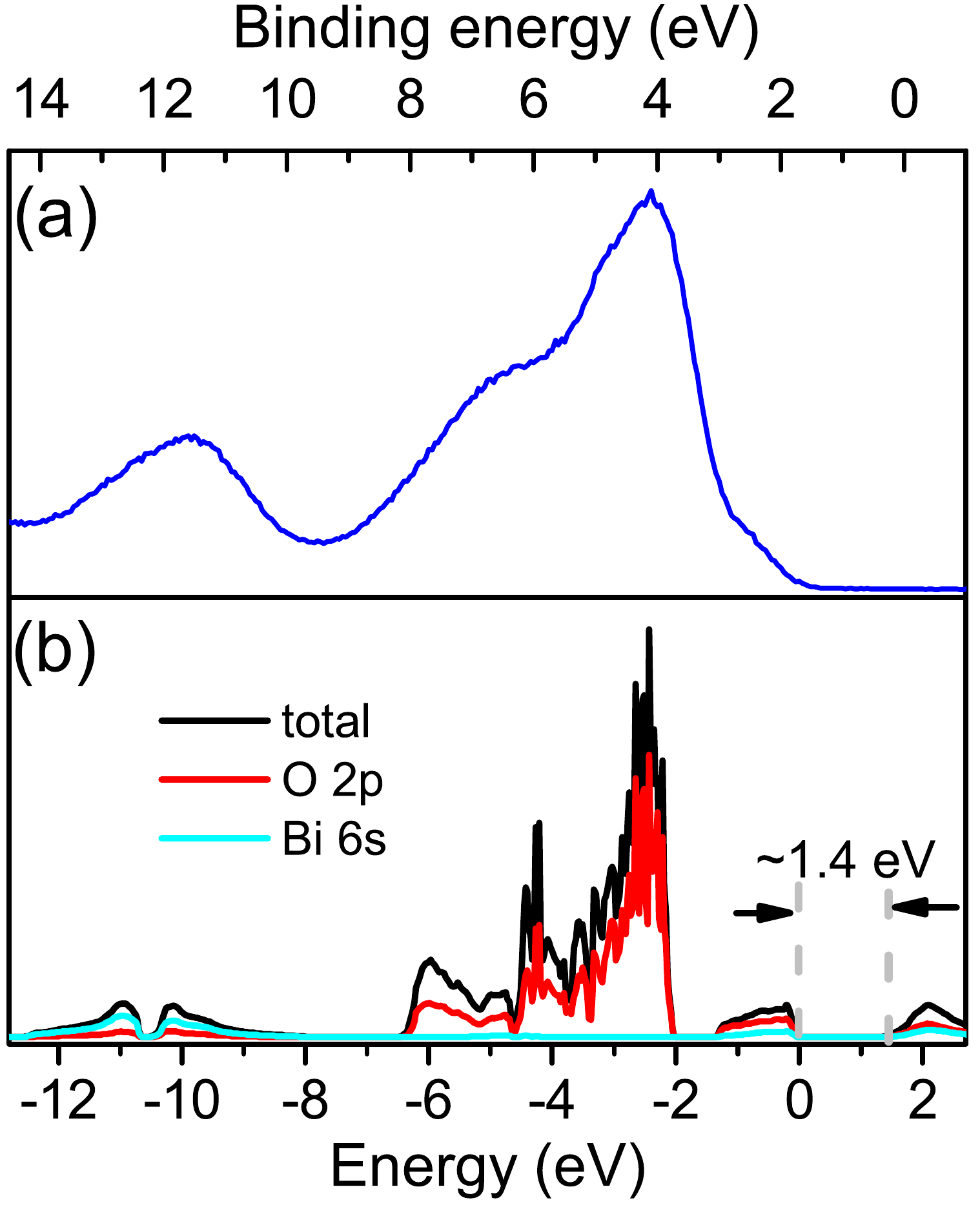}
	\caption{(a) Valence band information collected by XPS on the 20 UC SBO film grown on Nb-doped STO using the high-temperature codeposition method. (b) total DOS and partial DOS for Bi 6s and O 2p calculated with HSE hybrid functional DFT \cite{foyevtsova2015hybridization} and the $\sim$1.4~eV indirect bandgap was obtained. Since the Fermi level position located in SBO band gap is uncertain in DFT calculation, peaks at $\sim$-2~eV in theory and $\sim$4~eV in experiment are aligned for comparison.}
\end{figure}

The valence band measured by XPS on the 20~UC film surface is shown in Fig.~4(a). For comparison with the XPS result, as shown in Fig.~4(b), total and partial DOS are calculated using a $P2_1/n$ unit cell, a 6~$\times$~6~$\times$~5 grid of \textit{k} points and the Heyd-Scuseria-Ernzerhof (HSE) hybrid functional in Vienna \textit{ab initio} simulation package \cite{foyevtsova2015hybridization,kresse1996efficiency,paier2005perdew}. The spectral weight in Fig.~4(a) starts to increase at $\sim$1.4~eV, suggesting its $\sim$1.4~eV band gap and the film Fermi level is pinned closer to the conduction band in band gap \cite{foyevtsova2015hybridization}.  The peak with $\sim$12~eV binding energy in XPS [Fig.~4(a)] corresponds to the bonding state of Bi 6s and O 2p with the A$_{1g} $ symmetry. The spectral features between $\sim$10~eV and $\sim$2~eV are mainly derived from O 2p nonbonding states. The antibonding state in the expanded BiO$_6 $ octahedron just below the Fermi level shown in Fig.~4(b)  \cite{foyevtsova2015hybridization} is seen as a small shoulder at $\sim$2~eV binding energy in Fig.~4(a).

In conclusion, high-quality SBO films down to 1~UC are successfully prepared on SrO-terminated STO substrate. The successful fabrication of high-quality stoichiometric SBO film opens the possibility of manipulating the electronic states of perovskite bismuthate by interfacial engineering with or without strain applied by the substrates. In addition, low-energy electronic structures of relaxed SBO film show good agreement with theoretical calculation and supports the existence of strong hybridization of Bi 6s and O 2p in theoretical calculation that leads to its semiconducting nature.

\begin{acknowledgments}
The work was supported by Natural Sciences and Engineering Research Council of Canada (NSERC), Canada Foundation for Innovation (CFI), Canada First Research Excellence Fund (CFREF) and Canadian Institute for Advanced Research (CIFAR). Spectroscopy and part of film growth were performed at at the Surface Science Facility of the REIXS beamline in the Canadian Light Source, which is funded by the Canada Foundation for Innovation, NSERC, the National Research Council of Canada, the Canadian Institutes of Health Research, the Government of Saskatchewan, Western Economic Diversification Canada, and the University of Saskatchewan. The authors also acknowledge the financial support from MPI-CPfS (Dresden) for the experiment in the Canadian Light Source.
\end{acknowledgments}

\bibliography{ref}

\end{document}